\documentclass{appolb}
\usepackage{epsfig}

\begin{document}
\title{TRI$\mu$P - A New Facility to Investigate Fundamental Interactions with
   Optically Trapped Radioactive Atoms
\thanks{Presented at PAAT2002}%
}
\author{Klaus Jungmann
\footnote{representing work of the TRI$\mu$P group \cite{trimp_group} at KVI}
\address{Kernfysisch Versneller Instituut, Rijksuniversieit Groningen, Zernikelaan 25, NL 9747 AA Groningen}
}
\maketitle
\begin{abstract}
At the Kernfysisch Versneller Instituut (KVI) in Groningen, NL,
a new facility (TRI$\mu$P) is under development. It aims for
producing, slowing down and trapping of
radioactive isotopes in order to  perform accurate measurements
on fundamental symmetries and interactions.
A spectrum of radioactive nuclids
will be produced in direct, inverse kinematics of fragmentation
reactions using heavy ion beams from the superconducting
AGOR cyclotron.
The research programme pursued by the KVI group includes
precision studies of nuclear
$\beta$-decays through $\beta$--neutrino (recoil nucleus) momentum
correlations in weak decays and searches for permanent electric dipole
moments in heavy atomic systems. 
This offers  a large potential for discovering new physics or to limit
parameters in models beyond standard theory significantly. 
The scientific approach chosen in TRI$\mu$P can be regarded
as complementary to such high energy physics.
The facility in Groningen will be open for use by 
the worldwide community of scientists.
\end{abstract}
\PACS{12.15.Ji,12.10.Dm,29.30.Aj,32.80.Pj}
  
\section{Introduction - The Standard Theory}
Atomic physics has played an important and crucial role in the development of
modern physics. Precision measurements have lead quite often to unexpected
discoveries such as the existence of several isotopes for one element
or the discovery of anomalous magnetic moments. In these  cases small
and faint unexpected signals or tiny deviations from the standard 
theoretical treatment of atoms were observed.
They have every time eventually lead to a 
new picture of fundamental physics.

As an example, the manifestation of the deviation at the permille level of the
electron g-factor from the value 2 predicted in the Dirac theory
together with the observation of the Lamb shift in atoms have 
given rise to th best quantum field theory we know, 
Quantum Electrodynamics (QED) \cite{Kinoshita_90}. Today there is little serious doubt about 
the validity of QED in particular its underlying concepts. 
Only the sometimes very difficult calculations
present mathematical problems and discussions on the best suited
technical approaches occur. However, in all known cases
agreement has been reached at an admirable level of precision
between theory and experiments over 10 orders of magnitude in 
momentum transfer. 

QED is the U(1) part of the so called 
electroweak Standard Model (SM) which has SU(2)$\times$U(1) symmetry \cite{Langacker_95}. 
Together with the understood features
of strong interaction -- mainly known as Quantum Chromodynamics (QCD)--
it forms the standard theory. This is a powerful theoretical reference 
framework which allows to describe all observations in particle
physics until the turn of the millenium. Although the 
recently discovered, spectacular neutrino oscillations
are not fully provided with standard theory, they 
can, however, be  most probably  incorporated in a straight forward way
without any significant changes to the basic structure of this
theory building. The model uses 12 fundamental fermions, which are the
six quarks (u,d,s,c,b,t) and the six
leptons (e,$\mu$,$\tau$,$\nu_e$,$\nu_{\mu}$,$\nu_{\tau}$), as
building blocks of matter and 12 bosons ($\gamma$, W$^+$, W$^-$, Z$^0$,
8 gluons) as mediators of the forces.
It should be mentioned that through the inclusion
of QED in the electroweak Standard Model all atomic physics and its wide range of
applications in various different branches of sciences is
fully covered by the standard theory.

Despite its success, the Standard Model leaves many 
physical questions open. It provides an accurate description
of all experimental observations in particle physics.
However, often it also lacks any deep explanation for them.
Among the open standing problems are the number of three
particle generations, the masses of the fundamental fermions 
(quarks and leptons), the origin of parity violation in weak 
interactions or the dominance of matter over antimatter 
in the universe. In order to provide better explanations
for some or all of these questions 
many speculative models have been invited. They carry names
like Left-Right Symmetry, Technicolor, Supersymmetry and 
many more. String- and Brane- or even M-theory offer
even a coherent description which includes gravitation
and quantum mechanics under one umbrella.
For testing the predictions of these models, there
are in principle two different conceptual approaches.
One searches directly for new processes and particles.
This is typically done at high energy facilities.
The alternative approach is to search for
deviations in the behaviour of systems which 
can be described to very high precision within
standard theory. This method, for example, can exploit the fact 
that properties of atomic systems can be well calculated in 
the framework of QED. Precision measurements could reveal
small deviations which may originate from New Physics.

\begin{figure}[thb]
\label{FIG1}
 \unitlength 1.0cm
  \begin{picture}(15,10)  
   \vspace*{-2.0in}
   \hspace*{-0.in} 
   \centering{
   \psfig{file=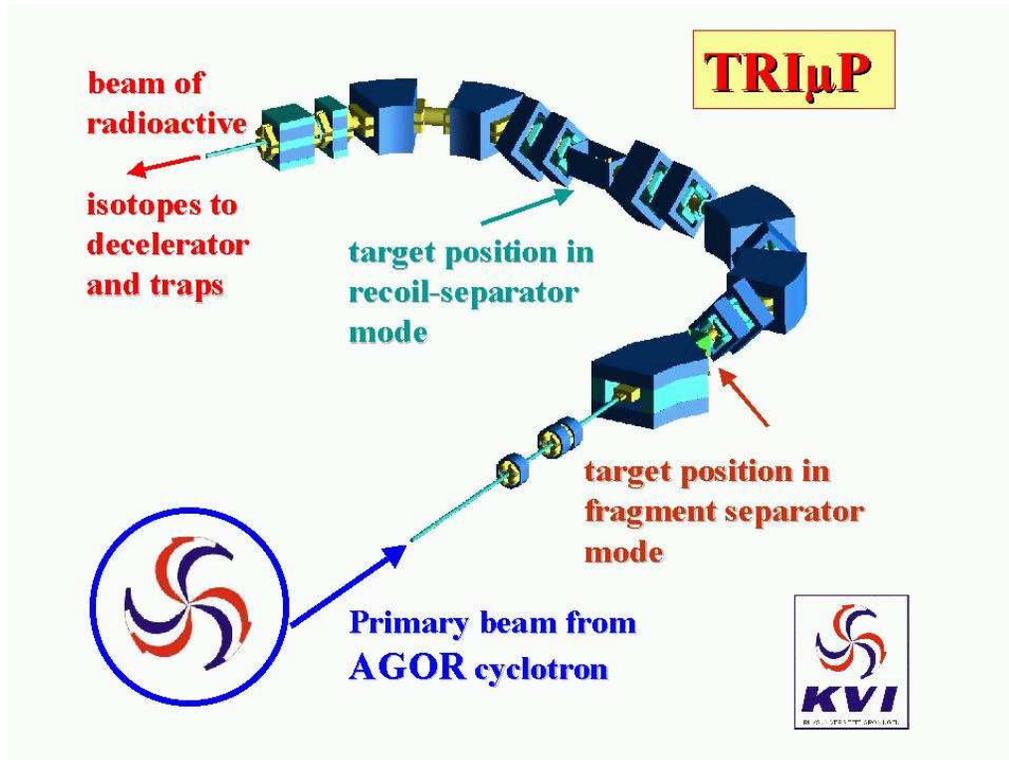,width=13.5cm,angle=0,clip=}
             }
  \end{picture}\par
  \caption[]{The TRI$\mu$P combined fragment and recoil separator.
It is designed to access a large variety of proton-rich isotopes.}
\end{figure}

\section{TRI$\mu$P - A New Facility}

At the Kernfysisch Versneller Insituut (KVI) in Groningen
a new facility - TRI$\mu$P ({\bf T}rapped {\bf R}adioactive {\bf I}sotopes: 
{\bf $\mu$}icro
laboratories for fundamental {\bf P}hysics) is at present being set up
\cite{Turkstra_99}.
It aims for producing
radioactive nuclids, slowing them down and trapping them
in atomic respective ion traps for precision experiments.

The isotopes are produced in direct, in inverse kinematics and
in fragmentation reactions. Heavy ion beams, which were 
accelerated in the the superconducting
AGOR cyclotron, are directed on fixed targets, the material of
which are chosen for optimal production rates. 

The created isotopes of interest need to be separated from
the primary beam and other reaction products. This can be achieved
in a combined fragment and (gas filled) recoil separator.
This device consists of mass and momentum selective ion optical
image system. The arrangement has two pairs of dipole magnets 
for the primary particle selection and quadrupoles
for accurate imaging (see figure \ref{FIG1}). The whole device foresees two 
possible target positions, one at its very entrance for fragmentation reactions
and another one between the two dipole pairs for inverse kinematics reactions.
Since the reaction products appear in a distribution
of charge states, gas filling of the separator is essential
for good imaging. The gas densities can be chosen 
such that electron capture and stripping result
in an effective charge  for the ions, which determines their trajectory
in the ion optical system \cite{Leino_97}.  

After the separator the nuclids of interest which typically have 
1 MeV/c momentum are 
slowed down in a (gas) moderator to energies in the eV range.
They are  further cooled by a buffer gas while being guided  
and radially confined by a radio frequency quadrupolar field 
in to a Paul Trap. The latter acts as a beam buncher.
After neutralization the atoms can be stored in atoms traps, e.g. a
magneto-optical trap. 

A user facility is created which is open to the worldwide
scientific community. TRI$\mu$P is jointly funded by FOM 
\footnote{Stichting voor Fundamenteel Onderzoek der Materie, 
Dutch funding agency} 
and the Rijksuniversiteit Groningen
in the framework of a managed programme. The time planning foresees
that the facility is set up by 2004 followed by an exploitation phase
until 2013. First physics experiments are expected in 2005.

\begin{figure}[thb]
\label{FIG1}
 \unitlength 1.0cm
  \begin{picture}(15,6.8)  
   \vspace*{2.0in}
   \hspace*{0.0in} 
   \centering{
   \psfig{file=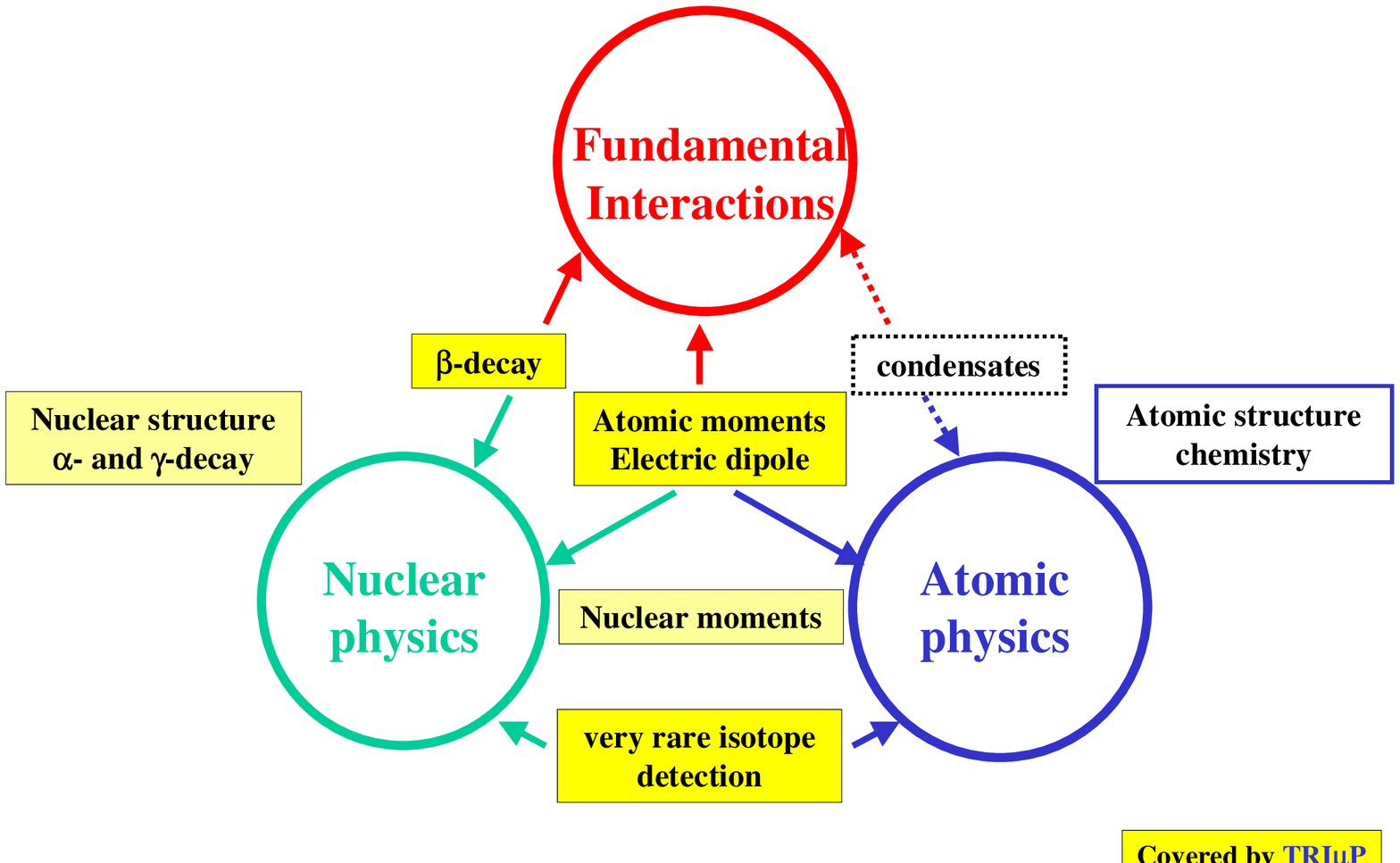,width=12cm,angle=0,clip=}
             }
  \end{picture}\par
  \caption[]{Atom and ion traps are employed in experiments
with their main goal in different fields of physics
and particularly in interdisciplinary research.}
\end{figure}

\section{Research at TRI$\mu$P}

Research using trapped radioactive atoms and ions 
covers a wide range of physics topics in atomic, nuclear and particle physics 
\cite{Schweikhard_00}
(see figure \ref{FIG2})).
The local group of researchers concentrates on two groups of experiments-
precision measurements of nuclear $\beta$-decays and searches for permanent
electric dipole moments in atoms.

\subsection{Precision Measurements of nuclear $\beta$-decays}

In standard theory the structure of weak
interactions is V-A, which means there are vector (V) and axial-vector (A) 
currents with opposite relative sign causing a left handed structure 
of the interaction and parity violation \cite{Herczeg_01}. 
Other possibilities like scalar, pseudo-scalar and tensor type 
interactions which might be possible would be clear 
signatures of new physics. So far they have been searched 
for without positive result. However, the bounds on parameters
are not very tight and leave room for various speculative possibilities. 
The double differential decay probability 
$ d^2W/d\Omega_e d\Omega_{\nu}$for a $\beta$-radioactive nucleus is
related to the electron and neutrino momenta $\vec{p}$ and $\vec{q}$ through
\begin{eqnarray}
\frac{d^2W}{d\Omega_e d\Omega_{\nu}} & \sim & 1 +  a ~\frac{\vec{p}\cdot\vec{q}}{E} 
+  b ~~\sqrt{1-(Z \alpha)^2}~~\frac{m_e}{E}            \\
& & + <\vec{J}>     \cdot \left[ A~~ \frac{\vec{p}}{E} + B~~\vec{q} + D~~\frac{\vec{p} \times ~\vec{q}}{E} \right]\\
& &+ <\vec{\sigma}> \cdot \left[ G~~ \frac{\vec{p}}{E} + Q~~\vec{J} + R~~ <\vec{J}> \times ~\frac{\vec{q}}{E} \right] 
\end{eqnarray}
where   $m_e$ is the $\beta$-particle mass,
        $E$ its energy,
        $\vec{\sigma}$ its spin,  and
        $\vec{J}$ is the spin of the decaying nucleus.
Among the coefficients D is of particular interest for
further restricting model parameters. It describes the correlation between 
the neutrino and $\beta$-particle momentum vectors for spin polarized nuclei 
and is parity violating in nature.

The coefficient R violates  parity and time reversal. However, it
relates to parameters in speculative models which are already well
constraint by searches for electric dipole moments (see below).
A direct measurement of the neutrino momentum is not efficiently possible.
Instead the recoiling nucleus can be detected instead and the neutrino 
momentum can be reconstructed using the  kinematics of the process.
Since the recoil nuclei have typical energies in the few 10 eV range,
precise measurements can only be performed, if the decaying isotopes are 
suspended using extreme shallow potential wells. Such exist, for example,
in magneto-optical traps, where many atomic species can be stored at 
temperatures below 1 mK. Since one needs to be able to trap the atoms 
optically and also the nuclear properties must be such that one has a 
rather clean transition, the isotopes of primary interest for
KVI experiments are $^{20}$Na, $^{21}$Na, $^{18}$Ne and $^{19}$Ne.
Optical trapping of both atomic Na and Ne is possible,
although, in the latter case trapping of metastable atoms is required \cite{Kuppens_02}.
In the case of Na successful recoil spectroscopy experiments
has been carried out in the framework of atomic charge transfer reactions \cite{Turkstra_02}.
For example, right-handed currents could give rise to deviations 
from standard theory predictions, which are searched for. Such
observations would be expected to shine light
into the mysteries behind parity violation in
weak interactions. 

\subsection{Permanent Electric Dipole Moments in Atoms}
 
A permanent electric dipole moment of any fundamental particle
violates both parity and time reversal symmetries \cite{Khrip_97}. 
With the assumption of 
CPT invariance the permanent dipole moment also violates CP.
The CP violation as it is known from the Kaon systems causes
through higher order loops permanent electric dipole moments
for all particles which are at least 4 orders of magnitude below the
present experimentally established limits. It should be noted
that the known sources of CP violation are not sufficient in 
Sakharov's model for the baryon asymmetry, i.e. the dominance of matter 
over antimatter in the universe \cite{Trodden_99}. New sources of CP violation 
would need to be discovered. Indeed, a large number of speculative models
foresees permanent electric dipole moments which could be as large as 
the present experimental limits just allow. 
Historically the non-observation of permanent electric dipole moments has
ruled out more speculative models than any other experimental
approach in all of particle physics \cite{Ramsey_99}. 

{
\begin{table}[bht]
\centering
\begin{tabular}{|c|rl|c|c|} \hline
& Present Limit & on              & Standard Model   & New Physics           \\
&  $|d|$        &                 & Prediction       & Limits               \\
& [10$^{-27}$&e\,cm]&  [10$^{-27}$e\,cm] & [10$^{-27}$e\,cm]                \\ \hline\hline
$e$        & $<1.6$          &{(90\% C.L.)}           & $\stackrel{ <}{\sim}10^{-11}$ 
                                                  & $\stackrel{ <}{\sim}1$   \\ \hline
$\mu$    & $<1.05\cdot10^9$  &{(95\% C.L.)}   & $\stackrel{ <}{\sim}10^{-8}$ 
                                                  & $\stackrel{ <}{\sim}200$ \\ \hline      
$\tau$   & $<3.1\cdot10^{11}$&{(95\% C.L.)} & $\stackrel{ <}{\sim}10^{-7}$
& $\stackrel{ <}{\sim}1700$                                                  \\ \hline 
$p$        & $-3.7\,(6.3)\cdot10^4$&              & $\sim10^{-4}$ &
$\stackrel{ <}{\sim}60$                                                      \\ \hline
$n$        &$<63$             &{(90\% C.L.)}              & $\sim10^{-4}$  &
$\stackrel{<}{\sim}60$                                                       \\ \hline
$^{199}$Hg&
$<0.21$                       &{(90\% C.L.)}              & $\sim10^{-6}$  &
$\stackrel{ <}{\sim}0.2$                                                     \\ \hline
\end{tabular}
\caption{Limits on Permanent Electric Dipole Moments $d$ for 
         electrons ($e$) \cite{Commins_94},
         muons ($\mu$)   \cite{Bailey_78},
         tauons ($\tau$) \cite{Aciari_98},
         protons ($p$)   \cite{Cho_89}, 
         neutrons ($n$)  \cite{Harris_99}, and the
         mercury atom ($^199$Hg)    \cite{Romalis_01}.
}     
\label{edm_limits}
\end{table}
}
Permanent electric dipole moments have been
searched for in various systems with different sensitivities 
(see table \ref{edm_limits}). In composed systems such as molecules
or atoms fundamental particle dipole moments can be enhanced
significantly \cite{Sandars_01}. Particularly in polarizable
systems there can exist large internal fields. 
Radium atoms in excited states are very interesting  
for electric dipole moment searches. 
Because of the rather close lying $7s7p^3P_1$ and $7s6d^3_2$ states
the a significant enhancement has been predicted  \cite{Dzuba_01}  
which gives an significant advantage over the mercury atom, the system 
which has given the best limits so far \cite{Romalis_01}.
From a technological point of view they well accessible 
spectroscopically and a variety of isotopes can be produced 
in fusion and evaporation or in fission reactions. 
The advantage of an accelerator based radium experiment is apparent,
because electric dipole moments require isotopes with spin
and all Ra isotopes with finite nuclear spin are relatively short-lived.

\subsection{Parity Violation in Atoms}

Precise measurements of weak charges in atomic parity violation experiments
and the precise electroweak parameter measurements at the LEP storage ring
facility at CERN
demonstrate together 
that the electroweak Standard Model covers interactions over 10 orders
of magnitude in momentum transfer to very high precision.
At present atomic parity experiments \cite{Haxton_01} may have found hints
to nuclear anapole moments. A severe limitation in the interpretation of 
the experiments arises from atomic structure calculations which are
possible at present to about 1\% accuracy. Weak interactions in atoms
scale with $Z^3$. The  Francium atom offers 18 times stronger weak
interaction effects and one expects calculations of the atomic structure
at the same level of accuracy as for the so far best system, the Cesium atom.
Since there are only rather short lived Francium isotopes,
such experiments are well suited to be carried out at accelerator sites.
Efforts exist at Legnaro, Italy, Stony Brook, USA, and Los Alamos, USA.
A second important issue is the distribution of neutrons in the nuclei
to which weak interactions are very sensitive. For Cesium a variation of
the neutron number means that radioactive isotopes will be involved.

\subsection{Applications}

For cold radioactive isotopes also a large variety of 
possibilities exists in research connecting to applied
sciences. For example one can imagine to use
cold polarized $\beta$-emitters which can be 
softly deposited on condensed matter surfaces.
This will allow to extend the method of
$\beta$-NMR, which is very successful in bulk material \cite{Fick_00},
to condensed matter surfaces.

\section{Conclusions}

The new facility TRI$\mu$P at KVI is expected to offer
new possibilities to study with high precision
fundamental interactions in physics and fundamental symmetries in
nature. The approach combines nuclear physics, atomic physics and 
particle physics in experimental techniques as well as in 
the conceptual approaches. The intense interaction
between theory and experiment will be necessary
for an optimal exploitation of the facility.
Photons, Atoms and all That will occupy an essential
and central part in this activity. Without 
the groundwork and the development of state of the art 
spectroscopy and atom manipulation techniques in the recent past
it would never be possible to achieve the desired precision,
which is necessary to arrive at relevant scientific conclusions.

\section{Acknowledgment}
This article is dedicated to Tomasz Dohnalik on the occasion of his 60$^{th}$
birthday.
The author feels privileged to know Tomasz as a scientist and 
as a friend. In periods when free communication and travel in Europe
was not the usual normality, as it naturally should be, our scientific 
relationship had important and unforgettable special human components as well. 
I am very grateful for Tomasz' friendship.  
The organizers of the PAAT2002 conference 
deserve thanks for providing a relaxed and
stimulating atmosphere.

\end{document}